\documentclass[prd,aps,showpacs,nofootinbib,eqsecnum,onecolumn,groupedaddress,amssymb]{revtex4}
\usepackage{amsmath}
\usepackage{amssymb}
\usepackage{amsfonts}
\usepackage{graphicx,bm}
\usepackage{dcolumn}
\usepackage{color,amsxtra}
\usepackage{epsf}
\usepackage{enumerate}
\usepackage{hhline}
\usepackage{array}
\usepackage{tabularx}
\usepackage{subfigure}
\usepackage{fancyhdr}
\usepackage{mathrsfs}

\newcommand{\be}{\begin{equation}}
\newcommand{\ee}{\end{equation}}
\newcommand{\bea}{\begin{eqnarray}}
\newcommand{\eea}{\end{eqnarray}}
\newcommand{\beaa}{\begin{eqnarray*}}
\newcommand{\eeaa}{\end{eqnarray*}}

\newcommand{\e}{\mathrm{e}}

\newcommand{\Eqn}[1]{&\hspace{-0.2em}#1\hspace{-0.2em}&}

\def\be{\begin{equation}}
\def\ee{\end{equation}}
\def\bea{\begin{eqnarray}}
\def\eea{\end{eqnarray}}

\def\e{\mathrm{e}}

\begin{document}

\title{Unification of Constant-roll Inflation and Dark Energy with Logarithmic $R^2$-corrected and Exponential $F(R)$ Gravity}

\author{
S.D. Odintsov$^{1,2}$, V.K. Oikonomou$^{3,4}$, L.~Sebastiani$^{5,6}$
}

\affiliation{ \\ \vspace*{2mm}
$^1$Consejo Superior de Investigaciones Cient\'{\i}ficas, ICE/CSIC-IEEC,
Campus UAB, Carrer de Can Magrans s/n, 08193
Bellaterra (Barcelona) Spain\\
$^2$Instituci\'{o} Catalana de Recerca i Estudis Avan\c{c}ats
(ICREA), Barcelona, Spain\\
$^3$International Laboratory for Theoretical Cosmology, Tomsk State
University of Control Systems and Radioelectronics (TUSUR), 634050
Tomsk, Russia\\
$^4$Tomsk State Pedagogical University, Tomsk, Russia\\
$^5$TIFPA (INFN)
Via Sommarive 14, 38123 Trento, Italy\\
$^6$Dipartimento di Fisica, Universit\`a di Trento, Via Sommarive 14, 38123 Trento, Italy }

\begin{abstract}
In this paper we investigate how to describe in a unified way a
constant-roll inflationary era with a dark energy era, by using the
theoretical framework of $F(R)$ gravity. To this end, we introduce
some classes of appropriately chosen $F(R)$ gravity models, and we
examine in detail how the unification of early and late-time
acceleration eras can be achieved. We study in detail the
inflationary era, and as we demonstrate it is possible to achieve a
viable inflationary era, for which the spectral index of primordial
curvature perturbations and the scalar-to-tensor ratio can be
compatible with the latest observational data. In addition, the
graceful exit issue is briefly discussed for a class of models.
Finally, we discuss the dark energy oscillations issue, and we
investigate which model from one of the classes we introduced, can
produce oscillations with the smallest amplitude.
\end{abstract}


\maketitle

\def\thesection{\Roman{section}}
\def\theequation{\Roman{section}.\arabic{equation}}

\section{Introduction}

Modified gravity in it's various forms, has a prominent role in
describing the Universe's evolution
\cite{reviews1,reviews2,reviews3,reviews4,reviews5}. Particularly, a
vast number of phenomena related to various stages of the Universe's
evolution can be explained by modified gravity theories, both at an
astrophysical level and also at a galactic level, see for example
the review \cite{reviews1}. Among the various theories of modified
gravity, $F(R)$ gravity is the most commonly used and, to our
opinion, the most appealing modified gravity theory, due to the
simplicity offered by $F(R)$ gravity descriptions, but also due to
the conceptual rigidity offered by this theory. Among other
applications, the $F(R)$ gravity theoretical framework, offers the
possibility of a unified description of the early-time and late-time
acceleration eras \cite{inlationDE}, see also Refs. \cite{nostriexp}
(for reviews on the unification of the early and late-time
acceleration, see \cite{reviews1,reviews2,reviews3}). This unified
description is one of the most demanding aims of a theoretical
description, and various approaches towards this issue have been
presented in the literature.

In the present work we aim to present some $F(R)$ gravity models
which can describe in a unified way a constant-roll inflationary era
with a late-time acceleration era. The constant-roll inflation
description, is an alternative approach to the standard slow-roll
inflationary era, and it's implications have recently been studied
in the context of scalar-tensor theories
\cite{Inoue:2001zt,Tsamis:2003px,Kinney:2005vj,Tzirakis:2007bf,
Namjoo:2012aa,Martin:2012pe,Motohashi:2014ppa,Cai:2016ngx,
Motohashi:2017aob,Hirano:2016gmv,Anguelova:2015dgt,Cook:2015hma,
Kumar:2015mfa,Odintsov:2017yud,Odintsov:2017qpp}, see also
\cite{Lin:2015fqa,Gao:2017uja,Gao:2017owg,Fei:2017fub} and also in
the context of $F(R)$ gravity
\cite{Nojiri:2017qvx,Motohashi:2017vdc,Oikre}. As it was shown in
\cite{Nojiri:2017qvx}, it is possible that non-viable $F(R)$ gravity
models in the context of slow-roll inflation, may become viable in
the context of constant-roll inflation. As we will demonstrate, in
the context of the constant-roll evolution, it is possible to obtain
observational indices that are compatible with the Planck
constraints. We shall use two well-known $F(R)$ gravity models,
namely an $R^2$-corrected logarithmic model and a curvature
corrected exponential model, and we explicitly calculate the
spectral index of primordial curvature perturbations and the
scalar-to-tensor ratio, and as we show, these are compatible with
the observational data. Also we shall briefly discuss some
interesting features of the reheating era corresponding to the
$R^2$-corrected logarithmic model. In addition, we analyze the
late-time evolution corresponding to both the aforementioned $F(R)$
gravity models, and it is noteworthy that these models provide a
unified description of the late and early-time acceleration era.
Finally, we analyze the dark energy oscillations for the
curvature-corrected exponential model, and we demonstrate that the
most phenomenologically appealing case corresponds to nearly $R^2$
curvature corrections.

This paper is organized as follows: In section II we introduce and
study two models of $F(R)$ gravity, and we investigate the
constant-roll inflation implications on the inflationary era. For
the $R^2$-corrected logarithmic model of $F(R)$ gravity we also
briefly discuss the reheating era implications. In section III, we
study the late-time evolution implications of the two models we
introduced in the previous section, and for the curvature corrected
exponential model, we investigate which curvature correction
produces less dark energy oscillations during the last stages of the
matter domination era and after the deceleration acceleration
transition. Finally, the conclusions follow in the end of the paper.

Before we start, let us briefly present the conventions we shall
assume. With regard to the geometric background, we assume that it
is a flat Friedmann-Robertson-Walker (FRW), with the line element
being,
\begin{equation}
\label{metricfrw} ds^2 = - dt^2 + a(t)^2 \sum_{i=1,2,3}
\left(dx^i\right)^2\, ,
\end{equation}
where $a(t)$ is the scale factor. In addition, we assume that the
metric connection is an affine, torsion-less, and metric-compatible
connection, the Levi-Civita connection. Also in the following, the
parameter $\kappa$ stands for $\kappa^2=16\pi/M_{\text{Pl}}^2$,
where $M_{\text{Pl}}$ is the Planck mass scale.

\section{The Inflationary Era with $R^2$-corrected Logarithmic and Exponential $F(R)$ Gravity}

\subsection{Model I of Inflation: $R^2$-corrected Logarithmic $F(R)$ gravity}

The first model we shall consider has the following action,
\begin{equation}
I=\int_\mathcal{M}d^4\sqrt{-g}\left[\frac{R}{\kappa^2}+\gamma(R)R^2+
f_{\text{DE}}(R)+\mathcal L_m\right]\,,\label{action}
\end{equation}
where $\mathcal M$ denotes the spacetime manifold, $g$ is the
determinant of the metric tensor $g_{\mu\nu}$, $\mathcal L_m$ is the
Lagrangian of the matter and radiation perfect fluids and $R$ is the
Ricci scalar. The corrections to the Hilbert-Einstein Lagrangian of
General Relativity (GR) assume the form of $F(R)$-gravity and are
represented by the higher curvature $R^2$-term for the early-time
inflation, and a function of the Ricci scalar, $f_\text{DE}(R)$, for
the dark energy sector. The running coefficient $\gamma(R)$ in front
of $R^2$ depends also on the Ricci scalar and has been introduced in
order for the graceful exit from inflation to be possible. When
$\gamma(R)=\gamma$ is constant, then the Starobinsky inflationary
scenario is obtained, where the early-time de Sitter expansion is
governed by an $R^2$ gravity.

The first Friedmann equation for the model at hand is the following,
\begin{eqnarray}
0&=&\frac{6H^2}{\kappa^2}-\gamma(R)\left[6R\dot H-12H\dot R\right]
+\gamma'(R)\left[
24H R\dot R-6R^2\left(H^2+\dot H\right)
\right]
+\gamma''(R)\left[
6H R^2\dot R
\right]+\nonumber\\
&&
f_\text{DE}
-(6H^2+6\dot H)f_\text{DE}'(R)+6H \dot f'_\text{DE}(R)-\rho_m\,,
\label{F1}
\end{eqnarray}
where the ``dot'' denotes the derivative with respect to the cosmic
time and the ``prime'' denotes differentiation with respect to the
Ricci scalar. The Ricci scalar reads,
\begin{equation}
R=12H^2+6\dot H\,,
\end{equation}
with $H=\dot a(t)/a(t)$ being the Hubble parameter. In the equation
above, $\rho_m$ denotes the energy density of matter which satisfies
the following conservation law,
\begin{equation}
\dot\rho_m+3H(\rho_m+p_m)=0\,,
\end{equation}
where $p_m$ is the matter pressure.

In order to reproduce the early-time acceleration we introduce the
following expression for  the function $\gamma(R)$,
\begin{equation}
\gamma(R)=\gamma_0\left(
1+\gamma_1\log\left[
\frac{R}{R_0}
\right]
\right)\,, \quad 0<\gamma_0\,,\gamma_1\,,
\label{gamma}
\end{equation}
where $R_0$ is the curvature of the Universe at the end of inflation
and $\gamma_0\,,\gamma_1$ are positive dimensional constants. Note
that logarithmic corrections of the form appearing in Eq.
(\ref{gamma}) are actually motivated by one-loop corrections to
coupling constants in multiplicatively renormalizable
higher-derivative quantum gravity, for a general review, see
\cite{Buchbinder:1992rb}. Also it has been demonstrated
\cite{Elizalde:2017mrn,Myrzakulov:2014hca}, that such an
$R^2$-corrected logarithmic gravity predicts viable inflation.
Furthermore, these terms may also find a theoretical explanation in
the framework of holographic renormalization group flow and the
corresponding cosmological implications, in the context these issue
were discussed in Refs. \cite{Cai:2010kp,Cai:2010zw}. In fact,  if
one incorporates in a gravitational system an holographic surface
term, located near to the Hubble horizon, one obtains higher
derivative corrections to the FRW field equations. Thus, the
solutions of these equations may be used to describe the early-time
acceleration and, eventually, by making use of two holographic
screens, the late-time accelerated expansion of our Universe today
may be realized.

Returning to the model at hand, it turns out that
$\gamma(R_0)=\gamma_0$. Since we would like to avoid the effects of
$R^2$-gravity in the limit of small curvature, we require that the
following general condition holds true,
\begin{equation}
\gamma_1\ll \frac{1}{\log\left[\frac{R_0}{4\Lambda}\right]}\ll 1\,,
\label{gamma1cond}
\end{equation}
where $R=4\Lambda$ is the curvature of the Universe when the dark
energy is dominant, and $\Lambda$ is the Cosmological constant. In
the following, we will assume that $f_\text{DE}(R)$ and $\mathcal
L_m$ in (\ref{action}) are negligible in the limit of high
curvatures. The de Sitter solution with constant curvature
$R_\text{dS}=12 H_\text{dS}$ follows from (\ref{F1}) and it reads,
\begin{equation}
H_\text{dS}^2\kappa^2=\frac{1}{12\gamma_0\gamma_1}\,,\quad
R_\text{dS}\kappa^2=\frac{1}{\gamma_0\gamma_1}\,.
\label{dS}
\end{equation}
In the case of a constant value of $\gamma(R)=\gamma_0$, namely for
$\gamma_1=0$, the de Sitter solution is obtained as an asymptotic
limit of the first Friedmann equation, when the $R^2$-term is
dominates the evolution. This is the so-called Starobinsky model,
where the Hilbert-Einstein term guarantees a graceful exit from the
accelerated phase. Since in the Starobinsky model $1/\kappa^2\ll R$,
we have a regime of super-Planckian curvature. Here, the $R^2$ term
has the same order of magnitude as the Hilbert-Einstein term during
the inflationary era. In this case, the running constant $\gamma(R)$
in (\ref{gamma}) with $\gamma_1\neq 0$, determines the value of the
de Sitter solution as in Eq. (\ref{dS}).

If we perturb the de Sitter solution as follows,
\begin{equation}
H=H_\text{dS}+\delta H(t)\,,\quad |\delta H(t)/H_\text{dS}|\ll 1\,,
\label{Hpert0}
\end{equation}
by keeping first order terms with respect to $\delta H(t)$, we
obtain from Eq. (\ref{F1}),
\begin{equation}
\frac{12 H_\text{dS}}{\kappa^2}
\left[
\left(
1-24H_\text{dS}^2\gamma_0\gamma_1\kappa^2
\right)\delta H(t)+3\gamma_0\kappa^2\left(
2+3\gamma_1+2\gamma_1\log\left[
\frac{R_\text{dS}}{R_0}
\right]
\right)(3H_\text{dS}\delta \dot H(t)+\delta \ddot H(t) )
\right]\simeq0\,.
\label{peq}
\end{equation}
In the limit $R_0\ll R_\text{dS}$ the solution of this equation
reads,
\begin{equation}
\delta H(t)\simeq h_\pm \text{e}^{\Delta_\pm t}\,,
\quad
\Delta_\pm=\frac{H_\text{dS}}{2}
\left(-3\pm
\frac{\sqrt{\log\left[
\frac{R_\text{dS}}{R_0}
\right]\left(16+9\log\left[
\frac{R_\text{dS}}{R_0}
\right]\right)}}{\log\left[
\frac{R_\text{dS}}{R_0}
\right]}\right)\,,
\end{equation}
where $h_\pm$ are constants depending on the sign of $\Delta_\pm$.
When we choose the plus sign, the solution diverges and the de
Sitter expansion is unstable. Thus, by Taylor expanding, the
divergent solution is given by,
\begin{equation}
\delta H(t)\simeq h_+\exp\left[\frac{4 H_\text{dS}t}{3\log[R_\text{dS}/R_0]}\right]\,,
\label{deltaH}
\end{equation}
and in effect, we obtain,
\begin{equation}
H\simeq H_\text{dS}\left(
1-h_0\text{e}^{\frac{H_\text{dS}(t-t_0)}{\mathcal N}}
\right)\,,
\label{Htinfl}
\end{equation}
where $t_0$ is the time at the end of inflation when $R\simeq R_0$
and also $h_0$, $R_0$ and $\mathcal{N}$ stand for,
\begin{equation}
h_0=\frac{(H_\text{dS}-H_0)}{H_\text{dS}}\,,\quad
\mathcal N=\frac{3}{4}\log\left[\frac{R_\text{dS}}{R_0}\right]\,,\quad R_0=12 H_0^2\,.
\label{def}
\end{equation}
In order to study the behavior of the solution during the exit from
inflation, we introduce the $e$-foldings number,
\begin{equation}
N=\log\left[
\frac{a(t_0)}{a(t)}
\right]\equiv\int^{t_0}_t H(t)dt\,.
\label{N}
\end{equation}
By using Eq. (\ref{Htinfl}) we have,
\begin{equation}
N  \simeq H_\text{dS}(t_0-t)
\,,\label{tN}
\end{equation}
where we have assumed that  $\mathcal N \ll H_\text{dS}(t-t_0)$, or
equivalently $\mathcal N\ll N$. Thus, the Hubble parameter may be
expressed as follows,
\begin{equation}
H\simeq H_\text{dS}\left(
1-h_0\text{e}^{-\frac{N}{\mathcal N}}
\right)\,.
\label{H}
\end{equation}
At the beginning of inflation we have $\mathcal N\ll N$ and $H\simeq
H_\text{dS}$, while at the end of the early-time acceleration, when
$N=0$, one recovers $H=H_0$.

During the quasi de Sitter expansion of inflation the Hubble
parameter slowly decreases. The slow-roll parameters are defined as
follows,
\begin{equation}
\epsilon=-\frac{\dot H}{H^2}=\frac{1}{H}\frac{d H}{dN}\,,\quad
-\eta=\beta=\frac{\ddot H}{2H\dot H}\,, \label{epsilonbeta}
\end{equation}
where we assumed that the constant-roll condition holds true. At the
beginning of the early-time acceleration the first slow-roll
parameter $\epsilon$ is small, in which case the slow-roll
approximation regime is realized. For the solution (\ref{H}) in the
limit $\mathcal N\ll N$, we get,
\begin{equation}
\epsilon\simeq
\frac{h_0 \text{e}^{\frac{H_\text{dS}(t-t_0)}{\mathcal N}}}{\mathcal N}
 =\frac{h_0 \text{e}^{-\frac{N}{\mathcal N}}}{\mathcal N}\,.
\label{epsilonvalue}
\end{equation}
On the other hand, for the $\beta$ parameter we obtain a constant
value, namely,
\begin{equation}
\beta=\frac{1}{2\mathcal N}\,.\label{beta}
\end{equation}
This means that the model at hand satisfies the condition for
constant-roll inflation. This fact has an important consequence on
the form of the spectral index of primordial curvature
perturbations, which will be independent from the total number of
the $e$-foldings during inflation. The inflationary paradigm offers
two important predictions about the inhomogeneities of our Universe
at a galactic scale. Particularly, the perturbations around the FRW
metric lead to a non-flat spectral index $n_s$ and also lead to a
non-zero scalar-to-tensor ratio $r$.

In the case of $F(R)$-gravity, the inflationary indices have the
following form,
\begin{equation}
(1-n_s)\simeq \frac{2\dot\epsilon}{H\epsilon}=-\frac{2}{\epsilon}\frac{d\epsilon}{ d N}\,,\quad
r\simeq 48\epsilon^2\,.
\end{equation}
By calculating these, we obtain,
\begin{equation}
(1-n_s)\simeq 4\beta-2\epsilon\simeq\frac{2}{\mathcal N}\,,\quad
r\simeq
48\frac{h_0^2 \text{e}^{-2\frac{N}{\mathcal N}}}{\mathcal N^2}\,.
\label{indexvalues}
\end{equation}
We can see that in the computation of the spectral index $n_s$ we
can omit the contribution of $\epsilon$ which tends to vanish for
$\mathcal N\ll N$ (like in the Einstein frame case of the
Starobinsky inflation). Since the constant-roll inflationary
condition is assumed, it turns out that this index is in fact
independent on the total $e$-foldings number.

The latest Planck data \cite{Planck} constrain the spectral index
and the scalar-to-tensor ratio as follows,
\begin{equation}
\label{planckdata} n_s=0.9644\pm 0.0049\, , \quad r<0.10\, .
\end{equation}
As a consequence, we must require $\mathcal N\simeq 60$ in order to
obtain a viable inflationary scenario. This means that at the
beginning of inflation we have $60\ll N$, a condition which solves
the problem of initial conditions of the Friedmann Universe model we
study.

We can compare this result with the one corresponding to the
ordinary Starobinsky model with $\gamma_1=0$. In this case, the
Hilbert-Einstein term of (\ref{F1}) evaluated on the background
solution (specifically, $6H_\text{dS}^2/\kappa^2$) can be inserted
in the perturbed Friedmann equation and we obtain instead of Eq.
(\ref{peq}), the following equation,
\begin{equation}
\delta \ddot H+3H_\text{dS}\delta \dot
H+\frac{H_\text{dS}}{12\gamma_0\kappa^2}\simeq 0\,,
\end{equation}
where, as we have discussed above, we have to consider
$1/\kappa^2\ll H_\text{dS}^2$. This equation implies the following,
\begin{equation}
\delta H(t)\simeq -\frac{1}{36\gamma_0\kappa^2}(t-t_\text{i})\,,
\end{equation}
where $t_\text{i}$ is appropriately chosen to be the time instance
at the beginning of inflation. Note that now we can eliminate the
non-linear term of the solution, such that $ \delta \ddot H(t)\simeq
0$ and the slow-roll parameter $\beta$ in (\ref{epsilonbeta}) is
approximately equal to zero. We also require that for $t=t_0$, when
inflation ends, the Hubble parameter vanishes, namely,
\begin{equation}
t_0\simeq 36\gamma_0\kappa^2 H_\text{dS}+t_\text{i}\,,
\end{equation}
such that,
\begin{equation}
H\simeq\frac{1}{36\gamma_0\kappa^2}\left(t_0-t\right)\,.
\label{HtStaro}
\end{equation}
For the $e$-foldings number (\ref{N}) we get,
\begin{equation}
N\simeq \frac{1}{72\gamma_0\kappa^2}\left(t_0-t\right)^2\,.
\end{equation}
As a consequence, the resulting Hubble rate solution has the
following form,
\begin{equation}
H\simeq\sqrt{\frac{N}{18\gamma_0\kappa^2}}\,.
\end{equation}
When $t=t_\text{i}$, the $e$-foldings number is given by
$N=18\gamma_0\kappa^2 H_\text{dS}^2$, and it is easy to verify that
$H=H_\text{dS}$.

The first slow-roll parameter $\epsilon$ for the Starobinsky
inflation is equal to,
\begin{equation}
\epsilon\simeq \frac{1}{2N}\,,\label{eS}
\end{equation}
and the spectral index and the scalar-to-tensor ratio are equal to,
\begin{equation}
(1-n_s)\simeq\frac{2}{N}\,,\quad r\simeq \frac{12}{N^2}\,.
\end{equation}
Therefore, we must require $N\simeq 60$ in order to have concordance
with the Planck data and therefore we need to specify the boundary
value of the Hubble parameter as follows $H_\text{dS}^2\simeq
60/(18\gamma_0\kappa^2)$. In the model at hand, the first slow-roll
parameter $\epsilon$ appearing in Eq. (\ref{epsilonvalue}) decreases
in exponential way, a behavior which is different in comparison to
Eq. (\ref{eS}). However, when $\mathcal N\simeq 60$ we obtain the
same spectral index with the Starobinsky inflation with $N\simeq
60$, but we need to point out that in our model, the total amount of
inflation measured by the $e$-foldings at the beginning of
inflation, can be larger than $\mathcal N\simeq 60$. This fact
contributes to easily suppress the value of the scalar-to-tensor
ratio in Eq. (\ref{indexvalues}), as it is strongly encouraged by
the BICEP2/Keck-Array data~\cite{BICEP2}.

By imposing $\mathcal N\simeq 60$ in Eq. (\ref{def}) we obtain,
\begin{equation}
R_\text{dS}\simeq R_0\text{e}^{80}\,,
\end{equation}
and the expansion curvature rate during inflation is defined in this
way. The characteristic curvature at the time of inflation is
$R_\text{dS}\simeq 10^{120} \Lambda$, in which case one has
$R_0\simeq 1.8\times 10^{85}\Lambda$ and from Eq. (\ref{gamma1cond})
we must require $\gamma_1\ll 0.005$. Finally, the relation between
$\gamma_0$ and $\gamma_1$ is fixed by Eq. (\ref{dS}) and we obtain,
\begin{equation}
\gamma_0\simeq\frac{\text{e}^{-80}}{\gamma_1  R_0\kappa^2}\,.
\end{equation}

\subsubsection{The Reheating Era}

At the end of the inflationary era, the model enters in a reheating
phase. We will still assume that the contribution of the term
$f_\text{DE}(R)$ in Eq. (\ref{F1}) is negligible, while the perfect
fluid matter content of the Universe (namely, the energy density
term $\rho_m$) has been shifted away, during the severe inflationary
accelerated expansion that the Universe experienced, and has to be
thermalized by the reheating process. During this phase, we will
assume that the radiation/matter fields are bosonic scalar fields,
which are described by the Klein-Gordon Lagrangian. By assuming that
$R\simeq R_0$ and by taking the limit (\ref{gamma1cond}),
Eq.~(\ref{F1}) reads,
\begin{equation}
\ddot H-\frac{\dot H^2}{2H}+\frac{1}{12\kappa^2\gamma_0}H\simeq -3H\dot H\,.
\label{eqr}
\end{equation}
The reheating solution takes place in an oscillatory phase, when the
following holds true,
\begin{equation}
|3H\dot H|\ll |\frac{\dot H^2}{2H}|\,.
\end{equation}
Thus, if one completely neglects the right hand side term of Eq.
(\ref{eqr}), we have $H(t)\propto\cos^2[\omega t]$ with
$\omega=1/\sqrt{24\gamma_0\kappa^2}$. In detail, one has to match
the solution (\ref{Htinfl}) with the new one in the form
$H(t)=f(t)\cos^2[\omega t]$, where $f(t)$ is a function of the time
such that $\dot f(t)^2/f(t)\simeq 0$ and $f(t)\dot f(t)\simeq 0$
(slowly-damping approximation). The form of the reheating
(oscillatory) solution is~\cite{Morris},
\begin{equation}
H(t)=\frac{1}{\frac{3}{\omega}+\frac{3}{4}(t-t_\text{r})+\frac{3}{8\omega}\sin\left[2\omega(t-t_\text{r})\right]}\cos^2\left[\omega(t-t_\text{r})\right]\,,
\label{genre}
\end{equation}
where $t_\text{r}$ is the time when the reheating starts. For
$t<t_\text{r}$, the solution is given by (\ref{Htinfl}) and
therefore, the transition to the reheating phase takes place when
$|\dot H^2/(2H)|\simeq |3H\dot H|$, namely,
\begin{equation}
t_\text{r}\simeq-\frac{\sqrt{\mathcal N}\sqrt{6\mathcal N(1-2h_0+h_0^2)+h_0}}{\sqrt{6}h_0 H_\text{dS}}+t_0\,,
\end{equation}
where we have expanded (\ref{Htinfl}) with respect to $H_\text{dS} |
t_\text{r}-t_0 |\ll 1$. Since $h_0\simeq 1$, if we use (\ref{tN}) we
get,
\begin{equation}
t_\text{r}\simeq\frac{N}{H_\text{dS}}-\frac{\sqrt{\mathcal N}}{\sqrt{6} H_\text{dS}}+t_\text{i}\simeq \frac{N}{H_\text{dS}}\,,
\end{equation}
with $\mathcal N\ll N$ being the total $e$-foldings of inflation.
By equating (\ref{Htinfl}) with (\ref{genre}) and by imposing
$h_0\simeq 1$, we can specify the frequency $\omega$ of the
reheating solution as follows,
\begin{equation}
\omega=\frac{3H_\text{dS}}{\sqrt{6\mathcal N}}=\frac{\sqrt{\beta}}{2\sqrt{\gamma_0\gamma_1\kappa^2}}\,,
\label{omegamio}
\end{equation}
where we have used Eq. (\ref{dS}) and we have introduced the
constant-roll parameter $\beta$ of Eq. (\ref{beta}).


For large values of $(t-t_\text{r})$ the Hubble parameter behaves as
follows,
\begin{equation}
H\simeq\frac{4}{3(t-t_\text{r})}\cos^2\left[
\omega(t-t_\text{r})
\right]\,.
\label{Hreheating}
\end{equation}
Since $<H>\simeq 2/(3(t-t_\text{r}))$, we get a matter dominated
cosmological evolution. By taking into account that $R\simeq 6\dot
H$, we obtain,
\begin{equation}
R\simeq -\frac{8\omega}{(t-t_\text{r})}\sin\left[2\omega(t-t_\text{r})\right]\,.
\label{Rreheating}
\end{equation}
The Lagrangian of a scalar bosonic field $\chi$ with mass $m_\chi$,
which is non-minimally coupled with gravity, is given by,
\begin{equation}
\mathcal L_\chi=-\frac{g^{\mu\nu}\partial_\mu\chi\partial_\nu\chi}{2}-\frac{m_\chi^2\chi^2}{2}-\frac{\xi R\chi^2}{2}\,,
\end{equation}
where $\xi$ is a coupling constant. Thus, the field equation is
derived as follows,
\begin{equation}
\Box\chi-m_\chi ^2\chi-\xi R\chi=0\,.
\end{equation}
As a consequence, the Fourier modes of the field
$\chi_k\equiv\chi_k(t)$ with momentum $k$ for a FRW spacetime,
satisfy the following differential equation,
\begin{equation}
\ddot\chi_k+3H\dot\chi_k+\left(m^2_\chi+\xi R\right)\chi_k=0\,.
\end{equation}
By introducing the conformal time $d\eta=dt/a(t)$ and also by
redefining $u_k= a(t)\chi_k$ we have,
\begin{equation}
\frac{d^2}{d\eta^2}u_k+m_\text{eff}^2 a(t)^2u_k=0\,,\quad
m^2_\text{eff}=\left[m_\chi^2+\left(\xi-\frac{1}{6}\right)R\right]\,.\label{u}
\end{equation}
As a result, the oscillations of the Ricci scalar change the
effective mass term $m_\text{eff}^2$ and the number of massive
particles $\chi_k$ increases with time. We note that, even in the
case of minimal coupling with gravity $\xi=0$, the effective mass
$m^2_\text{eff}$ still depends on the Ricci scalar $R$ and the
reheating mechanism takes place.

The energy density of radiation-ultra-relativistic matter is related
to the temperature as follows,
\begin{equation}
\rho=\gamma T^4\,,
\end{equation}
where $\gamma$ is some constant. Since the energy density is
connected to the square average of the Ricci scalar~\cite{Morris},
from Eq. (\ref{Rreheating}) we have $\rho\propto\omega^4$ and
therefore we obtain,
\begin{equation}
T\propto \omega=\frac{\sqrt{\beta}}{2\sqrt{\gamma_0\gamma_1\kappa^2}}\,,
\end{equation}
where we used (\ref{omegamio}).

In the Starobinsky inflationary scenario, the equation (\ref{eqr})
with solution (\ref{genre}) is still valid, but during the reheating
era we have,
\begin{equation}
t_\text{r}\simeq t_0+\sqrt{6\gamma_0\kappa^2}\simeq 36\gamma_0\kappa^2 H_\text{dS}
+\sqrt{6\gamma_0\kappa^2}+t_\text{i}\simeq 36\gamma_0\kappa^2 H_\text{dS}\,,
\end{equation}
or equivalently,
\begin{equation}
t_\text{r}\simeq \frac{2N}{H_\text{dS}}\,,
\end{equation}
with $N$ being the total number of $e$-foldings during the
inflationary era. In this case, the frequency $\omega$ in
(\ref{genre}) results to be,
\begin{equation}
\omega=\frac{1}{\sqrt{24\gamma_0\kappa^2}}\,.
\end{equation}
Thus, if we compare the reheating temperature of our model $T$ with
the reheating  temperature  corresponding to the Starobinsky
inflation $T_\text{St}$, we find,
\begin{equation}
\frac{T}{T_\text{infl}}=\sqrt{\frac{6\beta}{\gamma_1}}\,.
\end{equation}
Since the $\beta$ constant-roll parameter in (\ref{beta}) for
$\mathcal N\simeq 60$ is small, we see that the temperature of the
Universe after inflation predicted by our model is smaller respect
to the one of the standard $R^2$-scenario. This result shows that a
constant-roll inflationary era has important effects on the
reheating phase, as was also pointed out in Ref.~\cite{Oikre}.

\subsection{Model II of Inflation: Curvature-corrected Exponential $F(R)$ gravity with Constant-roll Evolution}

\subsubsection{Constant-roll Evolution in $F(R)$ Gravity}

In most theories of inflation in the context of $F(R)$ gravity, it
is common to adopt the slow-roll approximation, in order to produce
the right amount of inflation that may be compatible with the
observational data. In this section we shall deviate from the
standard slow-roll approach, and we shall assume that a
constant-roll era occurs during the inflationary era. The
constant-roll inflationary paradigm was firstly used in the context
of scalar-tensor theories
\cite{Inoue:2001zt,Tsamis:2003px,Kinney:2005vj,Tzirakis:2007bf,
Namjoo:2012aa,Martin:2012pe,Motohashi:2014ppa,Cai:2016ngx,Hirano:2016gmv,Anguelova:2015dgt,Cook:2015hma,
Kumar:2015mfa,Odintsov:2017yud,Odintsov:2017qpp}, see also
\cite{Lin:2015fqa,Gao:2017uja,Gao:2017owg,Fei:2017fub} and was
extended in the context of $F(R)$ gravity in
\cite{Nojiri:2017qvx,Motohashi:2017vdc,Oikre}. As it was shown in
\cite{Nojiri:2017qvx}, the most natural generalization of the
constant-roll condition in the Jordan frame is the following,
\begin{equation}
\label{constantrollcondition} \frac{\ddot{H}}{2H\dot{H}}\simeq
\beta\, ,
\end{equation}
where $\beta$ is some real parameter, which can be either positive
or negative. The condition (\ref{constantrollcondition}) is the most
natural generalization of the constant-roll condition used in
scalar-tensor approaches, which is,
\begin{equation}
\label{cr1} \frac{\ddot \phi}{ H \dot \phi} = \beta  \, ,
\end{equation}
since the condition (\ref{cr1}) is nothing else but the second
slow-roll index $\eta$, which in the most general case is equal to
$\eta \sim -\frac{\ddot{H}}{2H\dot{H}}$. We shall assume that the
theory is described by a vacuum $F(R)$ gravity and also that the
background is a flat FRW metric. Upon variation of the gravitational
$F(R)$ gravity action with respect to the metric, we get the
following equations of motion,
\begin{align}
\label{eqnmotion1}
3F_RH^2=& \frac{F_RR-F}{2}-3H\dot{F}_R \, , \\
\label{eqnmotion2} -2F_R\dot{H}=& \ddot{F}-H\dot{F} \, ,
\end{align}
where $F_R$ stands for $F_R=\frac{\partial F}{\partial R}$ and also
the ``dot'' denotes differentiation with respect to $t$. The
dynamics of inflation in the context of $F(R)$ gravity are governed
by four inflationary indices, $\epsilon_i$, $i=1,...4$, which are
defined as follows
\cite{Noh:2001ia,Hwang:2001qk,Hwang:2001pu,Nojiri:2016vhu,Odintsov:2016plw,Odintsov:2015gba,reviews1},
\begin{equation}
\label{slowrollgenerarlfrphi}
\epsilon_1=-\frac{\dot{H}}{H^2}\,,\quad \epsilon_2=0\, , \quad
\epsilon_3=\frac{\dot{F}_R}{2HF_R}\, ,\quad
\epsilon_4=\frac{\dot{E}}{2HE}\, ,
\end{equation}
with the function $E$ being equal to,
\begin{equation}
\label{epsilonfnction} E=\frac{3\dot{F}_R^2}{2\kappa^2}\, .
\end{equation}
Also for the calculation of the scalar-to-tensor ratio $r$, the
quantity $Q_s$ is needed, which is defined as follows,
\begin{equation}
\label{qsfunction} Q_s=\frac{E}{F_RH^2(1+\epsilon_3)^2}\, .
\end{equation}
The spectral index of primordial curvature perturbations $n_s$, in
the case that $\dot{\epsilon}_i\simeq 0$, is equal to
\cite{Noh:2001ia,Hwang:2001qk,Hwang:2001pu},
\begin{equation}
\label{spectralindex1} n_s=4-2\nu_s\, ,
\end{equation}
with $\nu_s$ being equal to,
\begin{equation}
\label{nus}
\nu_s=\sqrt{\frac{1}{4}+\frac{(1+\epsilon_1-\epsilon_3+\epsilon_4)
(2-\epsilon_3+\epsilon_4)}{(1-\epsilon_1)^2}}\, .
\end{equation}
The above relation is quite general and holds true not only in the
case that $\epsilon_i\ll 1$, but also when $\epsilon_i\sim
\mathcal{O}(1)$. With regard to the scalar-to-tensor ratio, in the
context of vacuum $F(R)$ gravity theories, it is defined as follows,
\begin{equation}
\label{scalartotensor1} r=\frac{8\kappa^2Q_s}{F_R}\, ,
\end{equation}
where the quantity $Q_s$ is given in Eq.~(\ref{qsfunction}) above,
and for the specific case of a vacuum $F(R)$ gravity, the
scalar-to-tensor ratio is equal to,
\begin{equation}
\label{scalartotensor2} r=\frac{48\epsilon_3^2}{(1+\epsilon_3)^2}\,
.
\end{equation}
The constant-roll condition (\ref{constantrollcondition}), affects
the inflationary indices of inflation $\epsilon_i$, $i=1,...,4$
appearing in Eq.  (\ref{slowrollgenerarlfrphi}), which can be
written as follows \cite{Nojiri:2017qvx},
\begin{equation}
\label{frgravityconstantroll}
\epsilon_1=-\frac{\dot{H}}{H^2}\,,\quad \epsilon_2=0\, , \quad
\epsilon_3=\frac{\dot{F}_{RR}}{2HF_R}\left(
24H\dot{H}+\ddot{H}\right)\, ,\quad
\epsilon_4=\frac{F_{RRR}}{HF_R}\dot{R}+\frac{\ddot{R}}{H\dot{R}}\, ,
\end{equation}
where $F_{RR}$ and $F_{RRR}$ stand for $F_{RR}=\frac{\partial^2
F}{\partial R^2}$ and $F_{RRR}$ $F_{RRR}=\frac{\partial^3
F}{\partial R^3}$ respectively.

It is conceivable that the inflationary dynamics crucially depend on
the functional form of the $F(R)$ gravity. We shall study a
curvature corrected exponential $F(R)$ gravity, which as we show,
perfectly describes the early-time acceleration era. Also as we show
in a later section, the late-time acceleration era is also
successfully described by this model.

\subsubsection{Constant-roll Evolution with Curvature-corrected
Exponential $F(R)$ Gravity}

\begin{figure}[h]
\centering
\includegraphics[width=18pc]{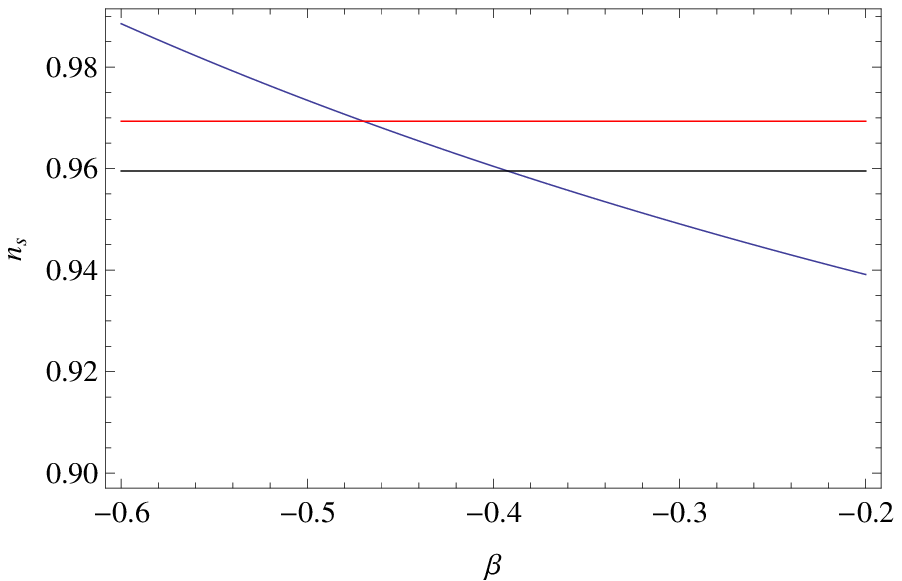}
\includegraphics[width=18pc]{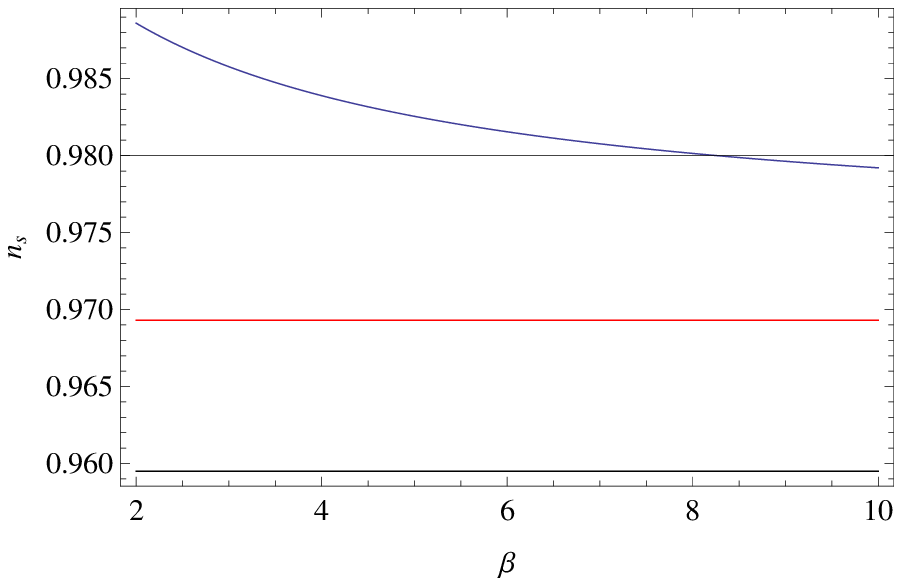}
\caption{The spectral index $n_s$ as a function of $\beta$, for
$n=1.5$ (blue curve, left plot) and for $n=2.1$ (blue curve, right
plot).}\label{plot1}
\end{figure}

\begin{figure}[h]
\centering
\includegraphics[width=18pc]{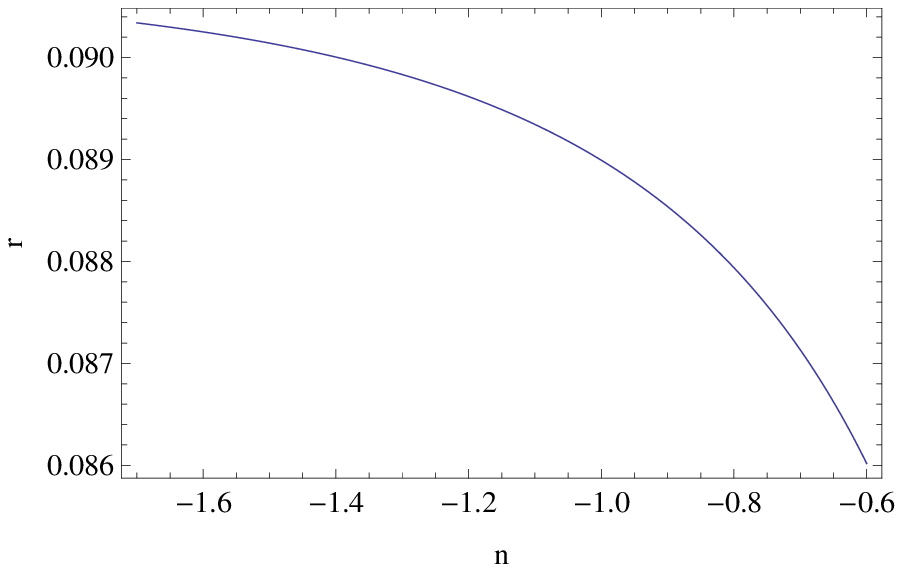}
\includegraphics[width=18pc]{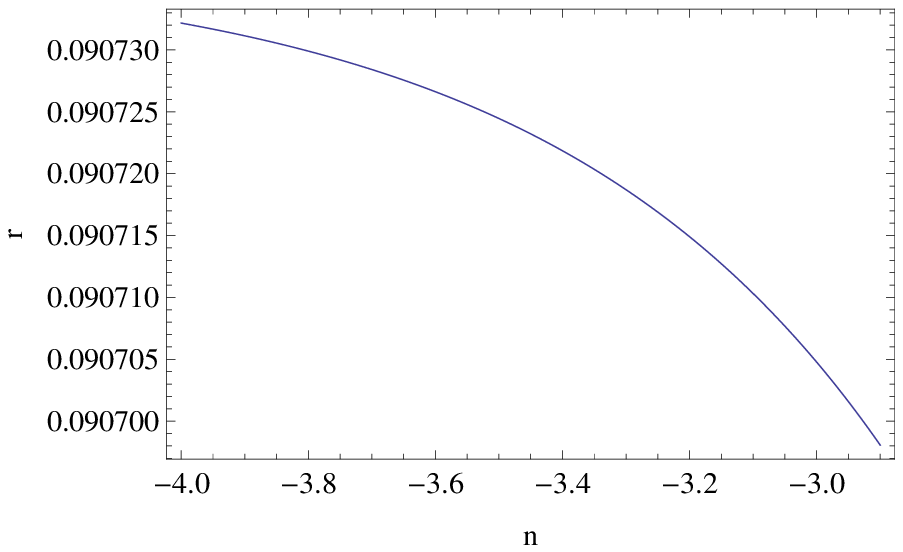}
\caption{The scalar-to-tensor ratio $r$, as a function of the
parameter $\beta$, for $n=1.5$ (left plot) and for $n=3$ (right
plot).}\label{plot2}
\end{figure}

The constant-roll condition utterly changes the viability of an
$F(R)$ gravity model, as it was shown in Ref. \cite{Nojiri:2017qvx}.
Particularly, it is possible that an $F(R)$ gravity model is not
compatible with the observational data, when this is studied in the
context of the slow-roll condition, however if the constant-roll
condition is assumed, the model may be compatible with the
observations, for a large range of values of the free parameters of
the model. In this section we shall present one model of this sort,
which is a curvature-corrected exponential model, with the
functional form of the $F(R)$ gravity being
\cite{oscillations1,oscillations2},
\begin{equation}
\label{expmodnocurvcorr} F(R)=R-2\Lambda \left (
1-\e^{\frac{R}{b\Lambda}} \right)-\tilde{\gamma}\Lambda
\left(\frac{R}{3m^2} \right )^{n} \, ,
\end{equation}
where $\Lambda=7.93m^2$ , $\tilde{\gamma}=1/1000$, $m=1.57\times
10^{-67}$eV, $b$ is an arbitrary parameter
\cite{oscillations1,oscillations2,Oikonomou:2014gsa} and $n$ is a
positive real parameter. This model has quite appealing inflationary
dynamics in the context of constant-roll inflation, as we now
evince. Consider the first equation of Eq. (\ref{eqnmotion1}), which
for the model (\ref{expmodnocurvcorr}) and by assuming that $H^2\gg
\dot{H}$ during the inflationary era, it can be approximated as
follows,
\begin{align}\label{tapprox1}
& 36 \gamma  n H(t)^{2 (n-1)+2}+\gamma  6^n H(t)^{2 n}-\gamma  6^n n
H(t)^{2 n}+\frac{6 \Lambda  H(t)^2 e^{-\frac{12
H(t)^2}{\text{R0}}}}{\text{R0}}+\Lambda  e^{-\frac{12
H(t)^2}{\text{R0}}}+3 H(t)^2-\Lambda \\ \notag & 864 \gamma  n^2
H(t)^{2 (n-2)+2} \dot{H}(t)-864 \gamma  n H(t)^{2 (n-2)+2}
\dot{H}(t)+\frac{144 \Lambda  H(t)^2 e^{-\frac{12
H(t)^2}{\text{R0}}} \dot{H}(t)}{\text{R0}^2}++\frac{6 \Lambda
e^{-\frac{12 H(t)^2}{\text{R0}}} \dot{H}(t)}{\text{R0}}\\ \notag &
216 \gamma  n^2 H(t)^{2 (n-2)+1} \ddot{H}(t)-216 \gamma  n H(t)^{2
(n-2)+1} \ddot{H}(t)++\frac{36 \Lambda  H(t) e^{-\frac{12
H(t)^2}{\text{R0}}} \ddot{H}(t)}{\text{R0}^2}=0\, ,
\end{align}
where we have set $R_0=b\Lambda$ and
$\gamma=\frac{\Lambda}{1000}\frac{1}{(3m)^n}$ for notational
simplicity. It is conceivable that the dynamical evolution is not
affected by the terms containing the exponentials, at least in the
era where the approximation $H^2\gg \dot{H}$ holds true. By using
the constant-roll condition (\ref{constantrollcondition}) and after
some algebraic manipulations, Eq. (\ref{tapprox1}) can be simplified
as follows,
\begin{equation}\label{basicqreqnconstroll}
432 \beta   n^2 \dot{H}(t)+864 n^2 \dot{H}(t)-432 \beta n
\dot{H}(t)-864  n \dot{H}(t)- 6^n n H(t)^2+36  n H(t)^2+ 6^n
H(t)^2=0\, .
\end{equation}
The differential equation (\ref{basicqreqnconstroll}) can be
analytically solved, with the solution being,
\begin{equation}\label{solutionbasicconst}
H(t)=\frac{432 (\beta +2) (1-n) n}{432 (\beta +2) C_1 (n-1)
n+\left(\left(6^n-36\right) n-6^n\right) t}\, ,
\end{equation}
with $C_1$ being an arbitrary integration constant which plays no
role in the dynamics of inflation, as we shall show. Also it is
notable that the parameter $\gamma$ appearing in Eq.
(\ref{tapprox1}), does not appear in the final differential equation
that governs the dynamical evolution. Having the Hubble rate at
hand, it is easy to calculate the slow-roll indices $\epsilon_i$,
$i=1,...4$, so by substituting Eq. (\ref{solutionbasicconst}) in Eq.
(\ref{frgravityconstantroll}), we obtain,
\begin{align}\label{constrollind}
\epsilon_1=\frac{6^n-\left(6^n-36\right) n}{432 (\beta +2) (n-1)
n},\,\,\,\epsilon_2=0,\,\,\,\epsilon_3=-\frac{-6^n n+36 n+6^n}{864
n},\,\,\,\epsilon_4=\frac{\left(\left(6^n-36\right) n-6^n\right) (-2
\beta +(\beta +2) n-1)}{432 (\beta +2) (n-1) n}\, ,
\end{align}
and by using these, the parameter $\nu_s$ defined in Eq.
(\ref{nus}), is equal to,
\begin{equation}\label{nusdef2}
\nu_s=\frac{1}{2} \sqrt{\frac{\left((\beta +2) \left(6^n+1260\right)
n^2-n \left(4 \beta  \left(6^n+297\right)+3
\left(6^n+852\right)\right)+(3 \beta +1) 6^n\right)^2}{\left(-432
(\beta +2) n^2+n \left(432 \beta -6^n+900\right)+6^n\right)^2}}\, ,
\end{equation}
and therefore in this case, the spectral index of primordial
curvature perturbations $n_s$ can be cast after some algebra in the
following form,
\begin{equation}\label{spectralindexcurbcorr}
n_s=4-\sqrt{\frac{\left(6^n (n-1) (-3 \beta +(\beta +2) n-1)+36 n
(-33 \beta +35 (\beta +2) n-71)\right)^2}{\left(36 n (-12 \beta +12
(\beta +2) n-25)+6^n (n-1)\right)^2}}\, .
\end{equation}
Accordingly, by using Eq. (\ref{scalartotensor2}), the
scalar-to-tensor ratio is found to be,
\begin{equation}\label{scalartotensorrationnew}
r=\frac{48 \left(6^n-\left(6^n-36\right)
n\right)^2}{\left(6^n-\left(6^n+828\right) n\right)^2}\, .
\end{equation}
It is noteworthy that both the spectral index and the
scalar-to-tensor ratio depend only on $\beta$ or $n$. Having the
final expressions for the spectral index and the scalar-to-tensor
ratio at hand, we shall investigate the parameter space in order to
see for which values of the free parameters, the compatibility with
the observational data can be achieved. The latest Planck data
\cite{Planck} constrain the spectral index and the scalar-to-tensor
ratio as in Eq. (\ref{planckdata}), so now we investigate which
values of the parameters $(n,\beta)$ may render the
curvature-corrected constant-roll model of Eq.
(\ref{expmodnocurvcorr}) compatible with the Planck data. A detailed
analysis reveals that there is a large range of parameter values
that may render the model compatible with the observations. For
example by choosing $(n,\beta)=(2.1,-8.7)$, the spectral index
becomes $n_s=0.966239$ and the corresponding scalar-to-tensor ratio
becomes $r=0.0119893$. Also for $(n,\beta)=(0.9,-1.08)$, the
spectral index becomes $n_s=0.96742$ and the corresponding
scalar-to-tensor ratio becomes $r=0.0936944$. Finally for
$(n,\beta)=(1.5,-0.4)$, the spectral index becomes $n_s=0.960444$
and the corresponding scalar-to-tensor ratio becomes $r=0.0669277$.
In Fig.~\ref{plot1} we plotted the functional dependence of the
spectral index as a function of $\beta$, for $n=1.5$ (left plot),
and for $n=2.1$ (right plot). In both cases, In the red and black
straight lines correspond to the Planck data allowed values for the
spectral index, namely, $n_s=0.9693$ and $n_s=0.9595$ respectively.
The same applies for the scalar-to-tensor ratio, and in
Fig.~\ref{plot2}, we plotted the functional dependence of the
scalar-to-tensor ratio as a function of $n$. In the left plot, the
parameter $n$ is assumed to be $n=1.5$, while in the right plot
$n=3$.
Concluding, we demonstrated that the curvature-corrected $F(R)$
gravity model of Eq. (\ref{expmodnocurvcorr}) in the context of the
constant-roll condition, provides an inflationary era which is
compatible with the latest Planck data. An intriguing feature of the
model (\ref{expmodnocurvcorr}), is that the parameter $n$ plays a
crucial role in the late-time era. As we show in a later section,
certain values of $n$ may produce an undesirable amount of dark
energy oscillations, so the values of $n$ we used into this section,
will be further constrained by the dark energy oscillations study.

Before we continue, it is worth discussing the resulting
scalar-to-tensor ratio as it appears in Eq.
(\ref{scalartotensorrationnew}). The resulting scalar-to-tensor
ratio is quite an important result, that might be tested in the next
generation of CMB experiments. So far, there have been several
ground based CMB projects under design or already on going, which
includes the CMB-S4 \cite{Abazajian:2016yjj}, BICEP3
\cite{Hui:2016pfm}, and AliCPT \cite{Cai:2016hqj}, and so on. It is
interesting to note that, the operation of all these ground based
experiments are able to cover the full CMB sky without the help of
satellites such as WMAP or Planck, and correspondingly, the
observational limit on the tensor-to-scalar ratio could be improved
to about two orders of magnitude higher in comparison to the current
one. This means that the parameter space of the models under
consideration would be much constrained in the near future.

Before we close this section, it is worth discussing in brief the
possibility of a graceful exit from the inflationary era, for the
model (\ref{expmodnocurvcorr}). In general, the graceful exit issue
can be a cumbersome task and also a conceptual challenge, especially
in the context of $F(R)$ gravity. In the literature, the graceful
exit from inflation instance is identified with the time instance
that the slow-roll approximation ceases to hold true, which occurs
when the first slow-roll index becomes of the order
$\mathcal{O}(1)$. However, in Ref. \cite{graceexit} an alternative
viewpoint was provided for the graceful exit from inflation issue.
Particularly, the graceful exit from inflation could be triggered by
the occurrence of growing curvature perturbations at some point
during the inflationary era, which may disturb the inflationary
attractor of the theory. In effect, the dynamical attractor becomes
unstable, and hence ceases to be the final attractor of the theory,
and the cosmological dynamical system is no longer described by this
attractor, and therefore the dynamical cosmological evolution of
inflation is interrupted. We believe that this way of thinking is
conceptually more accurate for the description of the graceful exit
from inflation.

So let us now investigate if the inflationary solution
(\ref{solutionbasicconst}) ceases to be the final attractor of the
cosmological system, and in order to see this, we linearly perturb
this solution, as follows,
\begin{equation}
\label{perturbationfromdesitter} H(t)=H_0(t)+\Delta H(t)\, ,
\end{equation}
where we identify $H_0(t)$ with the solution
(\ref{solutionbasicconst}). By inserting Eq.
(\ref{perturbationfromdesitter}) in the first equation of Eq.
(\ref{eqnmotion1}), and after some tedious algebraic manipulations,
we obtain the following differential equation at leading order,
\begin{align}\label{finalexitfgra}
& -47775744 \beta ^4 \gamma -381708288 \beta ^3 \gamma -1143621504
\beta ^2 \gamma -1522810368 \beta  \gamma -760389120 \gamma \\
\notag &-2592 (8 \beta ^2+31 \beta +30 ) \gamma  t^2 \Delta H'(t)
+144 t  (-12 (4608 \beta ^3 +27624 \beta ^2+55199 \beta +36766 )
\gamma ) \Delta H(t)=0\, ,
\end{align}
which can be solved and the solution is,
\begin{equation}\label{solutionforgracefulexit}
\Delta H(t)=-\frac{12 (\beta +2) (48 \beta +95) (96 \beta +193)}{(72
\beta  (128 \beta +511)+36721) t}\, .
\end{equation}
A simple analysis can easily reveal the behavior of the perturbation
$\Delta H (t)$, for the various values of the parameter $\beta$, and
in fact it can be shown that for $\beta<0$, the perturbations decay
as $t^{-1}$, while for $\beta>0$, the perturbations grow as the time
evolves to larger values. Thus, when $\beta>0$, the cosmological
solution $H_0(t)$ of Eq. (\ref{solutionbasicconst}) is unstable
towards linear perturbations, and therefore this can be viewed as a
strong indication that the graceful exit from inflation actually
happens in this case. However, the full study of this issue deserves
an article focusing on this, since there are various questions that
remain unanswered, for example, how many $e$-foldings occur before
the end of inflation, what happens when non-linear terms are taken
into account and so on. Some of these tasks may prove quite
difficult to address, since the lack of analyticity makes the
problem quite cumbersome. We aim though in a future work, to address
the aforementioned tasks by using an autonomous dynamical system
approach.

\section{Late-time Acceleration Era}

In this section we will investigate the behavior of the model I
appearing in Eq. (\ref{action}) during the late-time era. In order
to do so, we need to introduce the dark energy part in the action,
in terms of the function $f_\text{DE}(R)$ appearing in Eq.
(\ref{action}). We will use a modified version of exponential
gravity, namely,
\begin{equation}
f_\text{DE}(R)=-\frac{2\Lambda g(R)(1-\text{e}^{-b R/\Lambda})}{\kappa^2}\,,\quad 0<b\,,\label{fDE}
\end{equation}
where $b$ is a positive parameter and $\Lambda$ is the cosmological
constant. The function of the Ricci scalar $g(R)$ is necessary to
stabilize the theory at large redshifts and we shall assume that it
has the following form,
\begin{equation}
g(R)=\left[1-c\left(\frac{R}{4\Lambda}\right)\log\left[\frac{R}{4\Lambda}\right]\right]\,,\quad
0<c\,,\label{gR}
\end{equation}
where $c$ is a real and positive parameter. We need to note that,
the existence of a quintom scenario where the equation of state
(EoS) parameter evolves crossing the phantom divide line, is
strictly connected with the possibility of having a stable de Sitter
epoch at late times. This scenario is supported by cosmological
observations, which allow an oscillatory behavior of the dark energy
EoS parameter around the line of the phantom divide, although the
data are still far from being conclusive. In this respect, an
exhaustive review about the quintom scenario can be found in Ref.
\cite{Cai:2009zp}, where several successful examples of quintom
cosmological models and their corresponding observational
consequences were analyzed.

As a general feature of the model, we immediately see that, at
$R=0$, one has $f_\text{DE}(R)=0$ and we recover the Minkowski
spacetime solution of Special Relativity. When $4\Lambda\leq R$,
$f_\text{DE}(R)\simeq -2\Lambda/\kappa^2$ we obtain the standard
evolution of the $\Lambda$CDM model. Moreover, since $|f_\text{DE}
(R)| \sim 10^{-120} M_{Pl}^4$, we have that the modification of
gravity for the dark energy sector is completely negligible in the
high curvature limit of the inflationary era, where $R/\kappa^2\sim
M_{Pl}^4$.

We will now verify that with the modification we introduced in Eq.
(\ref{fDE}), it is possible to pass the cosmological and local
tests. In order to do so, we introduce the following form of the
$F(R)$ gravity,
\begin{equation}
F(R)\simeq R+\kappa^2 f_\text{DE}(R)\,,
\label{FRpr}
\end{equation}
in which case the trace of the field equations results to the
following equation,
\begin{equation}
3\Box F_R(R)+R F_R(R)-2F(R)=\kappa_0^2 T\,,\label{tracefield}
\end{equation}
where $T$ is the trace of the stress-energy tensor of the
matter-radiation perfect fluids that quantify the matter content of
the Universe. When $g(R)\simeq 1$, it is easy to see that the
following conditions hold true,
\begin{equation}
|F_R(R)-1|\ll 1\,,\quad 0<F_{RR}(R)\,,\quad\text{when}\quad 4\Lambda<R\,.\label{viableFR}
\end{equation}
The first condition is necessary in order to obtain the correct
value of the Newton constant and avoid anti-gravitational effects
during the matter, radiation and dark energy eras, while the second
condition guarantees the stability of the model with respect to the
matter perturbations~\cite{Faraoni2,Song,inlationDE}.

When $T=0$, which corresponds to the vacuum $F(R)$ gravity case, the
non-trivial solution of Eq. (\ref{tracefield}) is given by the de
Sitter spacetime with $R=4\Lambda$. This is a consequence of the
fact that $g(R)=1$ and $f_{\text{DE}}(R)\simeq -2\Lambda$ when
$R=4\Lambda$. The perturbations around the de Sitter solution are
governed by the equation,
\begin{equation}
\left(\delta \ddot R+3H \delta \dot R - m_\text{eff}^2\delta
R\right)\simeq 0\,,\quad m_\text{eff}^2=\frac{1}{3}
\left(\frac{F_R(R)}{F_{RR}(R)}-R \right)\,,\quad |\delta R/R|\ll
1\,.
\end{equation}
For our model, the effective mass $m_\text{eff}^2$ turns out to be
positive, rendering the solution stable. This means that the de
Sitter spacetime is a final attractor of the cosmological system for
large values of the cosmic time, when in the expanding Universe the
contents of matter and radiation vanish.

During the matter and radiation domination eras, the model we used
mimics an effective cosmological constant, if the function $g(R)$ in
Eq. (\ref{gR}) is close to unity, namely
\begin{equation}
c\ll \left[\left(\frac{R}{4\Lambda}\right)\log\left[\frac{R}{4\Lambda}\right]\right]^{-1}\,,\quad
4\Lambda\leq R\ll R_0\,,
\label{bcond}
\end{equation}
where recall that $R_0$ is the curvature of the Universe at the end
of the inflationary era. For example, if $c=10^{-5}$, we obtain
$f_\text{DE}\simeq 2\Lambda/\kappa^2$ up to the value $R\simeq
4\Lambda\times 10^4$. For larger values of the curvature, matter and
radiation dominate strongly the evolution. Moreover, the conditions
(\ref{viableFR}) read,
\begin{equation}
\left(F_R(R)-1\right)\simeq \frac{c}{2}\left[1+\log\left[\frac{R}{4\Lambda}\right]\right]\ll 1\,,\quad
0<F_{RR}(R)\simeq\frac{c}{2R}\,.
\end{equation}
The second condition always holds true for positive values of $c$,
but on the other hand, the first condition requires a careful
investigation in order to avoid anti-gravitational effects  in the
large curvature limit. For example, assume that $R_0\simeq 10^{86}
\Lambda$, then if $c\simeq 10^{-5}$, we obtain $(F_R(R)-1)\simeq
10^{-3}$ and it is obvious that this condition is satisfied.

\subsection{The Radiation, Matter Domination Eras and Transition to Late-time Acceleration Era}

In order to investigate the behavior of our model during radiation
and matter domination eras, but also during the transition to the
late-time era, and until present-time, we need to introduce the
following variable,
\begin{equation}
y_H \equiv\frac{\rho_{\mathrm{DE}}}{\rho_{\mathrm{m}(0)}}\equiv\frac{H(z)^2}{m^2}-(z+1)^3-\chi
(z+1)^{4}\,,
\label{y}
\end{equation}
which is known as the ``scaled dark energy''
\cite{SuperBamba,reviews1}. This variable encompasses the ratio
between the effective dark energy and the standard matter density,
evaluated at the present time, with the matter density defined as
follows,
\begin{equation}
\rho_{\text{m}(0)}=\frac{6m^2}{\kappa^2}\,,
\end{equation}
where $m$ is the mass scale associated with the Planck mass. In the
expression (\ref{y}), the variable $z=\left[1/a(t)-1\right]$ denotes
the redshift as usual, and also $\chi$ stands for
$\chi\equiv\rho_{\mathrm{r}(0)}/\rho_{\mathrm{m}(0)}$.

If one extends the expression appearing in Eq. (\ref{FRpr}) as
follows,
\begin{equation}
F(R)=\kappa_0^2\left[\frac{R}{\kappa^2}+\gamma(R)R^2+ f_{\text{DE}}(R)\right]\,,\label{FR}
\end{equation}
in order to include the entire form of the gravitational Lagrangian
(\ref{action}), it is possible to derive from the FRW field
equations the following equation~\cite{SuperBamba,reviews1},
\begin{equation}
\frac{d^2 y_H(z)}{d z^2}+J_1\frac{d y_H(z)}{d z}+J_2
y_H(z)+J_3=0\,,
\label{superEq}
\end{equation}
where the functions $J_i$, $i=1,2,3$ stand for,
\begin{eqnarray}
J_1 \Eqn{=} \frac{1}{(z+1)}\left[-3-\frac{1}{y_H+(z+1)^{3}+\chi (z+1)^{4}}\frac{1-F_R(R)}{6m^2
F_{RR}(R)}\right]\,, \nonumber
\\
J_2 \Eqn{=}
\frac{1}{(z+1)^2}\left[\frac{1}{y_H+(z+1)^{3}+\chi (z+1)^{4}}\frac{2-F_R(R)}{3m^2 F_{RR}(R)}\right]\,, \nonumber
\\
J_3 \Eqn{=}
-3 (z+1)
\nonumber \\
&&
-\frac{(1-F_R(R))((z+1)^{3}+2\chi (z+1)^{4})
+(R-F(R))/(3m^2)}{(z+1)^2(y_H+(z+1)^{3}+\chi
(z+1)^{4})}\frac{1}{6m^2
F_{RR}(R)}\,.\label{J3}
\end{eqnarray}
The Ricci scalar is easily derived from Eq. (\ref{y}) and it is
equal to,
\begin{equation}
R=3 m^2 \left[4y_H(z)-(z+1)\frac{d y_H(z)}{d z}+(z+1)^{3}\right]\,.
\label{RRR}
\end{equation}
At the late time regime, where $z\ll 1$, we can avoid the
contribution of the matter and radiation fluids, in which case, the
solution of Eq. (\ref{superEq}) reads,
\begin{equation}
y_H\simeq\frac{\Lambda}{3m^2}+y_0\text{Exp}\left[\pm i\sqrt{\frac{1}{\Lambda F_{RR}(4\Lambda)}-\frac{25}{4}}\log[z+1]
\right]\,,
\end{equation}
with $y_0$ being an integration constant.  Since for the exponential
gravity $\Lambda F_{RR}(4\Lambda)\ll1$, the argument of the square
root is positive, in effect, dark energy oscillates around the
phantom divide line $w=-1$. The frequency of the oscillation with
respect to $\log[z+1]$ is given by,
\begin{equation}
\nu=\frac{1}{2\pi}\sqrt{\frac{1}{\Lambda F_{RR}(4\Lambda)}-\frac{25}{4}}\,.
\end{equation}
Generally speaking, since $\Lambda F_{RR}(4\Lambda)\simeq
2b^2\exp[-4b]$, the oscillation frequency at late times does not
diverge. Such a problem may emerge in the high curvature regime,
that is for redshifts that satisfy $0\ll z$, when the contribution
of the dark energy in Eq. (\ref{RRR}) is negligible. In this case,
after the expanding Eqs. (\ref{superEq})--(\ref{J3}) with respect to
$y_H/(z+1)^3\ll 1$, we find the following solution in the vicinity
of a given redshift $z$ \cite{oscillations1,oscillations2},
\begin{equation}
y_\text{H}(z+\delta z)\simeq \frac{\Lambda}{3m^2}+y_0\text{Exp}\left[\pm i\nu \delta z\right]\,,\quad |\delta z/z|\ll 1\,,
\end{equation}
where $y_0$ is an integration constant. Here, the oscillation
frequency of the dark energy is equal to,
\begin{equation}
\nu\simeq \frac{1}{2\pi\sqrt{R F_{RR}(R)}(z+1)}\,.
\end{equation}
As a consequence, when $R F_{RR}(R)$ is very close to zero, like in
the case of pure (vacuum) exponential gravity, the effective dark
energy induced by modified gravity oscillates with high frequency
and diverges, rendering the theory unstable. However in our model,
due to the presence of the function $g(R)$ chosen as in Eq.
(\ref{gR}), one has,
\begin{equation}
\nu\simeq\frac{\sqrt{2/c}}{2\pi(z+1)}\,.
\label{nu}
\end{equation}
This means that, back into the past, during the radiation and matter
domination eras, the frequency of the effective dark energy
oscillations, tend to decrease and the theory is protected against
singularities.

\subsection{Dark Energy Era for the Curvature-corrected Exponential Model}

In this section we shall analyze certain features of the late-time
evolution corresponding to the curvature-corrected exponential
$F(R)$ gravity model of Eq. (\ref{expmodnocurvcorr}), having to do
with dark energy oscillations. It is known that certain viable
models of dark energy produce dark energy density oscillations
during the matter domination era, and the frequencies of these
oscillations may diverge
\cite{oscillations1,oscillations2,staroomega,oscillations3}. In the
modified gravity theories and especially in the $F(R)$ gravity
theories, fourth order derivatives appear in the theory, and hence
high-frequency oscillations occur and the background may oscillate
in a rapid way. In effect, perturbation theory may break down, since
non-linearities may occur. It is therefore vital for $F(R)$ gravity
theories to overcome this theoretical obstacle. In this section we
shall discuss this problem and by using some numerical analysis for
the curvature corrected exponential model, and we shall investigate
in which cases the dark energy oscillations may occur.

\begin{figure}[h]
\centering
\includegraphics[width=15pc]{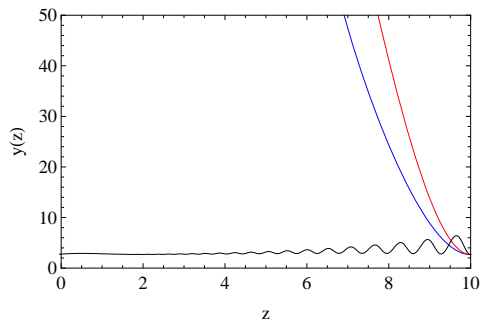}
\caption{Behavior of the scaled dark energy density
$y_H(z)=\frac{\rho_\mathrm{DE}}{\rho_m^{(0)}}$ as a function of the
redshift $z$, for $n=1.5$ (blue curve), $n=2.1$ (red curve) and
$n=0.9$ (black curve).}\label{ycomp}
\end{figure}

\begin{figure}[h]
\centering
\includegraphics[width=15pc]{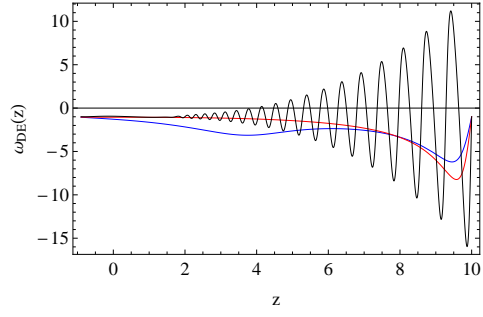}
\caption{The dark energy equation of state parameter
$\omega_\mathrm{DE}(z)$ as a function of the redshift $z$, for
$n=1.5$ (blue curve), $n=2.1$ (red curve) and $n=0.9$ (black
curve).}\label{omegadecomp}
\end{figure}

A convenient quantity that can easily quantify the dark energy
oscillations is the scaled dark energy
$y_H(z)=\frac{\rho_\mathrm{DE}}{\rho_m^{(0)}}$ appearing in Eq.
(\ref{superEq}). Actually, by solving the differential equation
(\ref{superEq}) numerically we will be able to find how the dark
energy oscillations behave at late-times, so for small redshifts.
For our numerical analysis we shall use the following initial
conditions
\cite{oscillations1,oscillations2,staroomega,oscillations3,Oikonomou:2014gsa},
\begin{equation}
\label{initialcond} y_H(z)\mid_{z=z_{f}}=\frac{\Lambda
}{3m^2}\left(1+\frac{z_{f}+1}{1000}\right)\, ,\quad
y_\mathrm{H}'(z)\mid_{z=z_{f}}=\frac{\Lambda}{3m^2} \frac{1}{1000}
\, ,
\end{equation}
with $z_f$ being $z_{f}=10$ and also $\Lambda $ is equal to $\Lambda
\simeq 11.89 $eV$^2$. In Fig.~\ref{ycomp} we have plotted the scaled
dark energy $y_H(z)=\frac{\rho_\mathrm{DE}}{\rho_m^{(0)}}$ for three
different values of the parameter $n$, namely $n=1.5$ (blue curve),
$n=2.1$ (red curve) and $n=0.9$ (black curve).
As we can see in Fig.~\ref{ycomp}, the scaled dark energy $y_H(z)$
for the cases $n=2.1$ and $n=1.5$ behaves in a very similar way.
Having the scaled dark energy at hand, we can easily investigate how
the dark energy oscillations behave by studying the dark energy
equation of state parameter
$\omega_\mathrm{DE}=P_\mathrm{DE}/\rho_\mathrm{DE}$, which in terms
of $y_H(z)$ is defined as follows,
\begin{equation}
\label{deeqnstateprm}
\omega_\mathrm{DE}(z)=-1+\frac{1}{3}(z+1)\frac{1}{y_H(z)}\frac{d
y_H(z)}{d z} \, .
\end{equation}
In Fig.~\ref{omegadecomp} we present the behavior of the dark energy
equation of state parameter for $n=1.5$ (blue curve), $n=2.1$ (red
curve) and $n=0.9$ (black curve).
As it can be seen in Fig.~\ref{omegadecomp}, the oscillations have
smaller amplitudes for $n=2.1$, and these start during the matter
domination era. In all cases, the dark energy oscillations stop to
occur at small redshifts. It is noteworthy that the values of $n$
around $n\sim 2$, always provide the smaller amplitudes for the
oscillations.

In order to have a clear picture for the physical behavior of the
cosmological system, we shall investigate the behavior of the total
equation of state parameter $\omega_\mathrm{eff}(z)$, as a function
of the redshift $z$, which in terms of $y_H(z)$ is defined as
follows,
\begin{equation}
\label{effeqnofstateform} \omega_\mathrm{eff}(z)
=-1+\frac{2(z+1)}{3H(z)}\frac{d H(z)}{d z}\, .
\end{equation}
\begin{figure}[h]
\centering
\includegraphics[width=15pc]{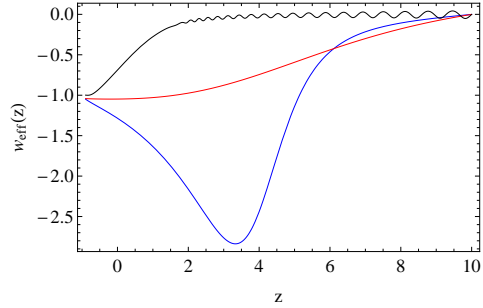}
\caption{The effective total equation of state parameter
$\omega_\mathrm{eff}(z)$ as a function of the redshift $z$, for
$n=1.5$ (blue curve), $n=2.1$ (red curve) and $n=0.9$ (black
curve).}\label{weffcomp}
\end{figure}
In Fig.~(\ref{weffcomp}) we plot the behavior of the total equation
of state parameter as a function of the redshift $z$, for $n=1.5$
(blue curve), $n=2.1$ (red curve) and $n=0.9$ (black curve). As it
can be seen, the case $n=2.1$ has the mildest oscillatory behavior.
It is noteworthy that in all cases, for small redshifts the phantom
divide line $w=-1$ is not crossed.

\begin{figure}[h]
\centering
\includegraphics[width=15pc]{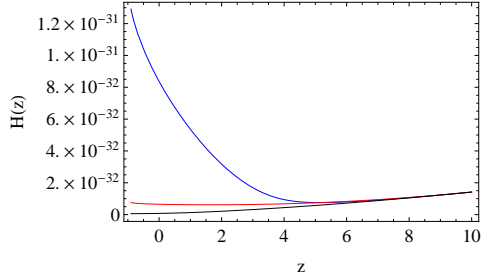}
\caption{Comparison of the Hubble parameter $H(z)$ as a function of
$z$ for $n=1.5$ (blue curve), $n=2.1$ (red curve) and $n=0.9$ (black
curve).}\label{hubcurv}
\end{figure}

Finally, in Fig.~\ref{hubcurv} we plot the Hubble rate as a function
of the redshift for $n=1.5$ (blue curve), $n=2.1$ (red curve) and
$n=0.9$ (black curve). Note that in terms of the scaled dark energy
parameter $y_H(z)$, the Hubble rate is defined as follows,
\begin{equation}
\label{hubblepar} H(z)=\sqrt{m^2y_H(z)+g(a(z))+\chi (z+1)^{4}}\, ,
\end{equation}
The analysis of the dark energy equation of state parameter and also
of the total equation of state parameter, indicates strongly that
the exponential model with nearly $R^2$ curvature corrections, has
quite appealing properties, since the dark energy oscillations are
less pronounced.

\subsection{Dark Energy Oscillations for the Model I}

Now let us investigate the dark energy oscillations issue for the
model I appearing in Eqs. (\ref{action}) and (\ref{gamma}). We
assume the following choices-conventions for the parameters,
\begin{equation}
\kappa^2=\frac{16\pi}{M_{Pl}^2}\,,
\quad\gamma_0=\frac{\text{e}^{-80}}{\gamma_1
R_0\kappa^2}\,,\quad\gamma_1=10^{-4}\,,\quad R_0=1.8\times
10^{85}\Lambda\,,\label{cond1}
\end{equation}
where,
\begin{equation}
M_{Pl}^2=1.2 \times 10^{28} \text{eV}^2 \,,\quad
\Lambda=1.1895\times 10^{-67}\text{eV}^2\,.
\end{equation}
The second condition in Eq. (\ref{cond1}) leads to a realistic de
Sitter curvature for the early-time acceleration, which is
$R_\text{dS}\simeq 10^{120}\Lambda$. Moreover, the third condition
in Eq. (\ref{cond1}) ensures that the high curvature corrections of
the model I disappear after the inflation, when $R<R_0$.

The constant parameters of the function $f_\text{DE}(R)$ in Eqs.
(\ref{fDE})--(\ref{gR}) are chosen as follows,
\begin{equation}
b=\frac{1}{2}\,,\quad c=10^{-5}\,.
\end{equation}
In this way, we obtain an optimal reproduction of the $\Lambda$CDM
model, and the effects of dark energy remain negligible during the
early and mid stages of the matter and radiation eras.

Now we need to fix the boundary conditions of our cosmological
dynamical system at large redshift $z=z_\text{max}$. They can be
inferred from the form of $\rho_\text{DE}$ in Eq. (\ref{y}) for the
case of $F(R)$-modified gravity, namely,
\begin{equation}
\rho_\text{DE}= \frac{1}{\kappa_0^2 F_R(R)}\left[ (R
F_R(R)-F(R))-6H\dot F_R(R) \right]\,.
\end{equation}
When $\Lambda\ll R\ll R_0$ we obtain,
\begin{equation}
y_H(z)\simeq\left(\frac{\Lambda}{3m^2}\right)\left(g(R)-6H^2
g_{RR}(R)(z+1) R\right]\,,
\end{equation}
where $R\equiv R(z)$ and $H\equiv H(z)$ are functions of the
redshift. At large redshift, during the matter era, we have to take
$R=3m^2(z+1)^3$ and  $H=m (z+1)^{3/2}$ and the boundary conditions
of the system are given by,
\begin{eqnarray}
y_H(z_\text{max})&=&\left(\frac{\Lambda}{3m^2}\right)\left[g(R_\text{max})-54 m^4(z_\text{max}+1)^6 g_{RR}(R_\text{max}) \right]\,,\nonumber\\
\frac{d y_H}{d z}(z_\text{max})&=& 3\Lambda(z+1)^2\left[
g_R(R_\text{max})-6 R_\text{max}^2 g_{RRR}(R_\text{max})-12
R_\text{max} g_{RR}(R_\text{max}) \right]\,,
\end{eqnarray}
where,
\begin{equation}
R_\text{max}=3m^2(z_\text{max}+1)^3\,.
\end{equation}
For $z_\text{max}=10$, in which case $\chi(z_\text{max}+1)\simeq
0.00341\ll 1$, and we effectively are in a matter dominated
Universe, we obtain,
\begin{equation}
y_H(z_\text{max})=2.1818\,,\quad \frac{d y_H}{d
z}(z_\text{max})=-2.6\times 10^{-5}\,,\quad z_\text{max}=10\,.
\end{equation}
These values can be compared with the corresponding ones for the
$\Lambda$CDM model, where $y_H$ is a constant, namely $
y_H=\Lambda/(3m^2)=2.17857 $. We argue that our model is extremely
close to the $\Lambda$CDM model at very high redshift. Here we
recall that the first observed galaxies correspond to a redshift $z
\simeq 6$.

\begin{figure}[!h]
\begin{center}
\includegraphics[angle=0, width=0.65\textwidth]{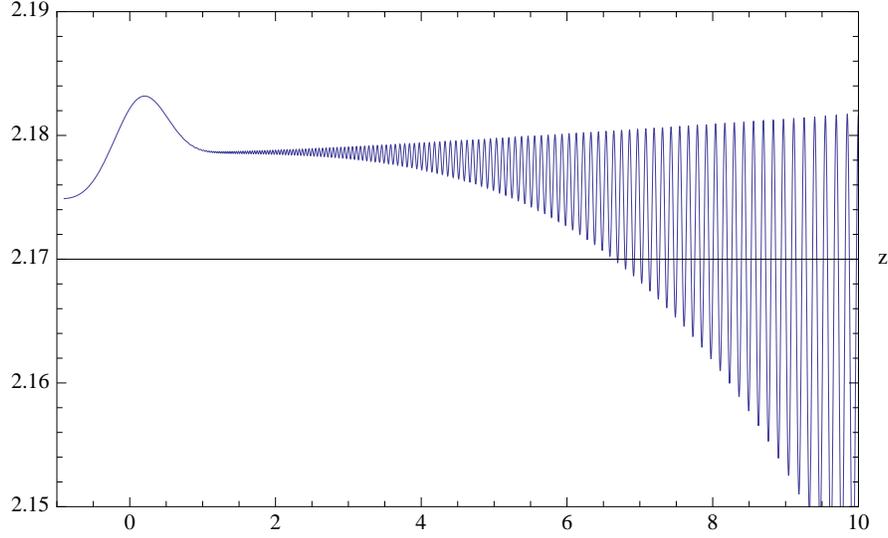}
\end{center}
\caption{Plot of $y_H(z)$ for $-1<z<10$.\label{Fig1}}
\end{figure}

\begin{figure}[!h]
\begin{center}
\includegraphics[angle=0, width=0.65\textwidth]{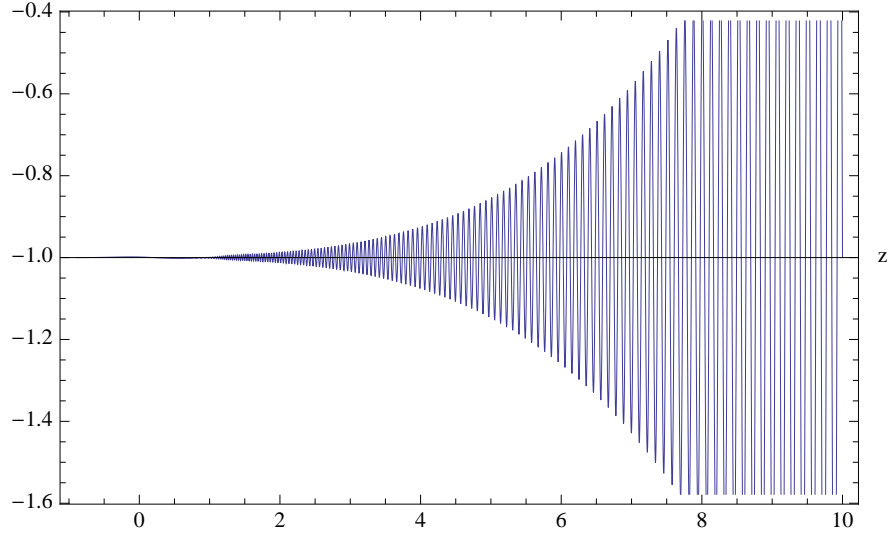}
\end{center}
\caption{Plot of $\omega_\text{DE}(z)$ for $-1<z<10$.\label{Fig2}}
\end{figure}

Finally, the contributions of matter and radiation are determined by
the values of $m^2$ and $\chi$ in (\ref{y}). The cosmological data
indicate that,
\begin{equation}
m^2\simeq 1.82 \times 10^{-67}\text{eV}^2\,,\quad \chi\simeq 3.1
\times 10^{-4}\,.
\end{equation}
By solving numerically the differential equation for $y_H(z)$ in the
redshift range, $-1<z<z_\text{max}$ we obtained the results which
appear in Fig.~(\ref{Fig1}) and Fig.~(\ref{Fig2}), where we plotted
the behavior of $y_H(z)$ and $\omega_\text{DE}(z)$ as a function of
the redshift, for $-1<z<10$.

\begin{figure}[!h]
\begin{center}
\includegraphics[angle=0, width=0.65\textwidth]{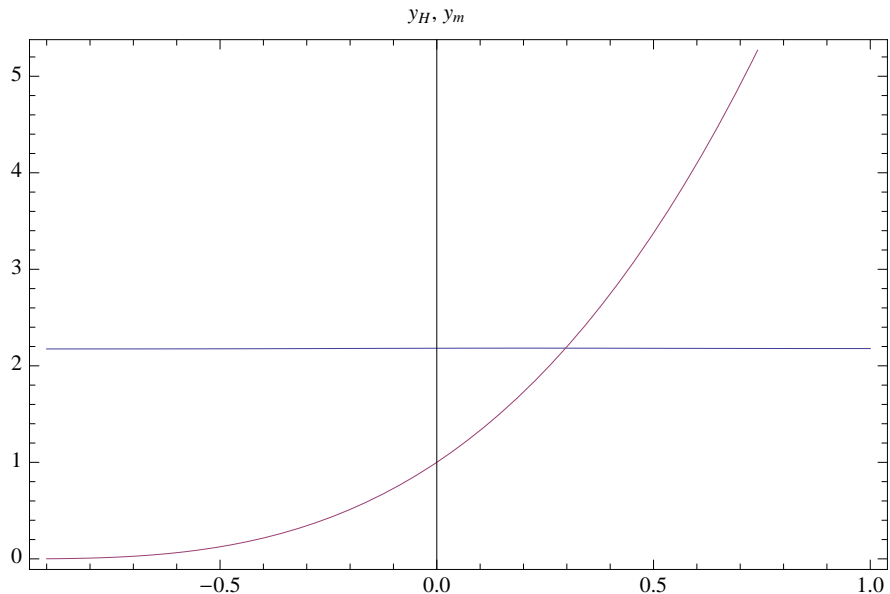}
\end{center}
\caption{Plot of $y_H(z)$ (blue line) and $y_m(z)$ (pink line) for
$-1<z<1$.\label{Fig3}}
\end{figure}

Despite of the fact that at high redshifts, the amplitude of the
oscillations of the effective EoS parameter around the phantom
divide line gradually grows, we see that their frequency decreases
and thus, singularities are avoided.
In order to measure the matter energy density $\rho_\text{m}(z)$ at
a given redshift, we introduce the parameter $y_m(z)$ as
\begin{equation}
 y_m(z)=\frac{\rho_\text{m}(z)}{\rho_{\text{m}(0)}}\equiv (z+1)^3\,.
\end{equation}
In Fig.~(\ref{Fig3}) the plot of $y_H(z)$ is compared with the
graphic of $y_m(z)$ for $-1<z<1$. We see that $y_H(z)$ is nearly
constant and it is dominant over $y_m(z)$,  for $z<0.4$, a feature
that is in full agreement with the $\Lambda$CDM description.
The $\Omega_\text{DE}(z)$ parameter,
\begin{equation}
\Omega_\text{DE}(z)\equiv\frac{\rho_\text{DE}}{\rho_\text{eff}}=\frac{y_H(z)}{y_H(z)+(z+1)^3+\chi(z+1)^4}\,,
\label{OMEGADE}
\end{equation}
is frequently used to express the ratio between the dark energy
density $\rho_\text{DE}$ and the effective energy density
$\rho_\text{eff}$ of our FRW Universe. Thus, by extrapolating
$y_H(z)$ at the current redshift $z=0$, from Eqs. (\ref{OMEGADE}),
we obtain,
\begin{equation}
\Omega_\text{DE}(z=0)=0.685683\,,\quad\omega_\text{DE}(z=0)=-0.998561\,.
\end{equation}
The latest cosmological data~\cite{Planck} indicate that,
$\Omega_\text{DE}(z=0)=0.685\pm0.013$ and
$\omega_\text{DE}(z=0)=-1.006\pm0.045$. Thus, our model fits the
observational data at present time.

Hence, we have shown that the model I can mimic the late-time
features of the $\Lambda$CDM Model. The additional degree of freedom
introduced by the $F(R)$-gravity, leads to an oscillatory behavior
during the last stages of the matter domination era and the early
stages of the dark energy era. However, the theory is stable and
protected against singularities. We need to stress that in the
simulation presented in this section, we have considered the whole
form of the gravitational Lagrangian in (\ref{action}). Thus, we
confirm that the higher curvature corrections for inflation are
negligible in the limit of small curvatures.

\section*{Conclusions}

In this paper we demonstrated that it is possible to have a unified
description of a constant-roll inflationary era with the dark energy
era, by using two $F(R)$ gravity models. Particularly, we used a
$R^2$-corrected logarithmic $F(R)$ gravity model and a
curvature-corrected exponential $F(R)$ gravity model. With regard to
the inflationary era, we demonstrated that it is possible to obtain
observational indices compatible with the latest Planck data, in the
context of constant-roll inflation. Particularly, the constant-roll
condition broadens the parameter space, and this feature makes it
more easy to have compatibility with the observational constraints.

In addition, we studied the late-time evolution of the $F(R)$
gravity models. A noticeable feature related with the exponential
model is that the dark energy oscillations have quite small
amplitude when the curvature corrections are nearly of the $R^2$
form.

It would be interesting to analyze the possibility of having a
unified description of constant-roll inflation with a dark energy
era, in the context of other modified gravities, like for example
Gauss-Bonnet or $F(T)$ gravities, where it is possible to unify
inflation with dark energy, see Ref. \cite{reviews1} for a
comprehensive review. This issue deserves a study which we defer in
a future work.

With regard to the graceful exit from inflation issue, which we
briefly addressed for one of the inflationary models we presented in
this paper, it should be noted that it deserves a more focused
analysis. With the approach we adopted in this work, we provided
hints that the graceful exit may occur as a result of growing
curvature perturbations, and with exit from inflation we mean that
an inflationary attractor solution may become unstable as the time
evolves. In a more focused work, one should in principle construct
an autonomous system of differential equations, and prove
numerically that the inflationary attractors are indeed unstable
stationary points of the dynamical system. Work is in progress
towards the aforementioned line of research.

Finally, when gravitational systems different from the Einstein
gravity are considered, we can find new cosmological solutions with
a large variety of new features. In this work, we have seen that a
constant roll inflation can be obtained by introducing higher
derivative corrections in the action, while a quintom scenario for
the dark energy can be obtained at late times, in the context of
$F(R)$-gravity. A modified theory of gravity may also allow for an
alternative description with respect to the Big Bang theory as in
the bounce scenario, where a cosmological contraction is followed by
an expansion at a finite time. The idea that instead from an initial
singularity the universe has emerged from a cosmological bounce has
been largely analyzed in the literature (see for example Refs.
\cite{Cai:2007qw,Cai:2011tc,Cai:2012va,Odintsov:2016tar,Odintsov:2015ynk,Haro:2015oqa}
or Ref. \cite{Cai:2015emx} for bounce cosmology description in the
context of teleparallel gravity).

\section*{Acknowledgments}

This work is supported by MINECO (Spain), project FIS2013-44881,
FIS2016-76363-P and by CSIC I-LINK1019 Project (S.D.O) and by
Ministry of Education and Science of Russia Project No. 3.1386.2017
(S.D.O and V.K.O).

\end{document}